# High-efficiency electro-optic modulator on thin-film lithium niobate with high-permittivity cladding

**NUO CHEN, KANGPING LOU, YALONG YU, XUANJIAN HE, TAO CHU**[*]

[1] *College of Information Science and Electronic Engineering, Zhejiang University, Hangzhou, 310027, China*
[*]*chutao@zju.edu.cn*

**Abstract**: Thin-film lithium niobate is a promising platform owing to its large electro-optic coefficients and low propagation loss. However, the large footprints of devices limit their application in large-scale integrated optical systems. A crucial challenge is how to maintain the performance advantage given the design space restrictions in this situation. This article proposes and demonstrates a high-efficiency lithium niobate electro-optic (EO) modulator with high-permittivity cladding to improve the electric field strength in waveguides and its overlap with optical fields while maintaining low optical loss and broad bandwidth. The proposed modulator exhibits considerable improvement, featuring a low half-wave voltage–length product of 1.41 V·cm, a low excess loss of ~0.5 dB, and a broad 3 dB EO bandwidth of more than 40 GHz. This modulation efficiency is the highest reported for a broadband lithium niobate modulator so far. The design scheme of using high-permittivity cladding may provide a promising solution for improving the integration of photonic devices on the thin-film lithium niobate platform and these devices may serve as fundamental components in large-scale photonic integrated circuits in the future.

## 1. Introduction

Lithium niobate (LiNbO3, LN) has been acknowledged as a dominant photonic material owing to its superior and versatile properties, such as a wide transparency window (∼0.35–5 µm) and strong electro-optic ($\gamma_{33}$=31 pm / V at 633 nm), ferroelectric, and piezoelectric coefficients.[1-3] In recent decades, there has been an increasing need for photonic integrated circuits to have lower power consumption and device footprints due to the increasing complexity of optical systems. The availability of wafer-scale and high-quality thin-film lithium niobate (TFLN) on insulators has garnered widespread interest in integrated optics[4-7] because TFLN not only inherits the excellent physical properties of LN, but also achieves stronger optical field confinement and improves the element integration density.[8-10]

As crucial components of photonic integrated circuits, electro-optic (EO) modulators are widely used in telecommunications, data communications, optical sensors, and quantum optics, among other applications.[11-16] TFLN modulators have experienced rapid development in recent years,[17–22] and numerous high-efficiency modulators have been demonstrated.[23–30]

Liu et al. increased the EO modulation efficiency to a voltage–length product of 1.75 V·cm using a shallowly etched lithium niobate waveguide.[29] Owing to the disparity between the dielectric constants of lithium niobate and silica, the electrical field primarily affected the LN core through the slab. The shallowly etched waveguide increased the proportion of the waveguide slab, enhancing the electric field density. However, the modulation efficiency achievable using this method had an upper limit owing to the restriction of light-propagation loss arising from electrode absorption. Another representative study demonstrated a new electrode extending from the sides to the top of the waveguide, creating a much stronger electric field within the narrow gap. The modulation efficiency was increased to 0.64 V·cm, the highest on record to date.[30] Nevertheless, this irregular electrode could have a considerable influence on microwave attenuation, with a measured EO bandwidth of approximately 3 GHz.

It is evident that the voltage–length product, bandwidth, and optical loss in the aforementioned reports continue to sustain critical trade-offs constrained by light-metal absorption and inefficient microwave signal delivery.[7] How to improve the modulation efficiency while preserving other favorable performance measures remains to be investigated.

This article proposes and demonstrates an LNOI high-efficiency EO modulator with composite cladding. The design using high-dielectric-constant cladding alters the electric field intensity distribution, which contributes to considerable improvements in the electric field strength in waveguides and enhances the overlap between the electric and optical fields. Additionally, the devices preserve low optical loss and high bandwidth simultaneously. That is, the proposed solution goes beyond the above constraints, introducing new variables to improve efficiency without affecting other performance measures. The proposed modulator exhibited a low excess loss of ~0.5 dB and a low voltage-length product of 1.41 V·cm. The 3 dB EO bandwidth of the modulator was measured to be greater than 40 GHz. This work provides a general solution for improving the integration of photonic devices on the TFLN platform with high-permittivity cladding.

## 2. Principle and Design

The complicated electric field distribution inside the phase shifter of the modulator is simplified to the electric field issue inside parallel plate capacitor, as shown in **Figure 1**a. A parallel dielectric capacitor consists of two parallel metallic conducting plates with a dielectric medium ($\varepsilon$), silicon dioxide ($SiO_2$), and LN filled in between.

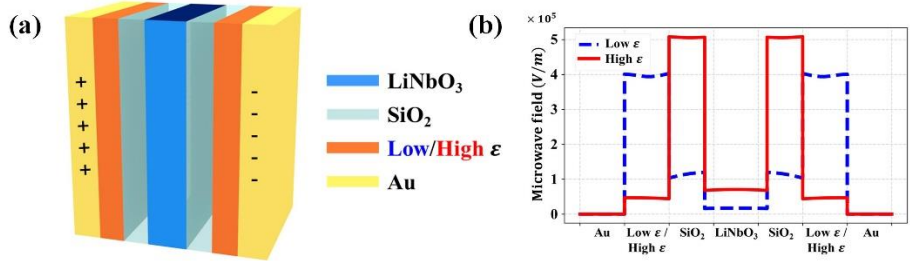

**Figure 1.** (a) Schematic diagram of the parallel plates. (b) Simulated microwave field distribution of electrode plates for low $\varepsilon$ (blue line) and high $\varepsilon$ (red line) media. The electric field inside LN has been increased by more than four times.

According to Gauss's law, at the interface of a medium with different $\varepsilon$, the electric field strength satisfies Maxwell's equation:

$$D = \varepsilon_1 E_1 = \varepsilon_2 E_2 \tag{1}$$

Substituting into the interior of the electrode plates yields

$$\varepsilon E = \varepsilon_{SiO_2} E_{SiO_2} = \varepsilon_{LN} E_{LN} \tag{2}$$

where $\varepsilon$, $\varepsilon_{SiO_2}$, and $\varepsilon_{LN}$ denote the relative permittivities of the medium, $SiO_2$, and LN, respectively, and $E$, $E_{SiO_2}$, and $E_{LN}$ denote the electric field strengths within them. The electric field strength inside LN can be improved by replacing the external medium with a larger dielectric constant. Low and high $\varepsilon$ values of approximately 1 and 44, respectively, are substituted into the calculation, and the electric field distribution diagram inside the electrode plate is shown in **Figure 1**b. The electric field inside LN has been increased by more than four times.

We introduced this idea into the modulator, whose structure is shown in **Figure 2.** It is designed on an LNOI wafer with a 600-nm-thick X-cut LN layer and a 4.7-µm-thick SiO$_2$ layer. The Mach–Zehnder interferometer (MZI) modulator consists of waveguide phase shifters and multimode interferometers. Gold traveling-wave electrodes are laterally spaced in a ground–signal–ground configuration. The composite claddings consist of a SiO$_2$ layer and an upper high-dielectric-constant material layer, hereinafter referred to as the upper cladding, as depicted in **Figure 2**b. The basic structural parameters are the waveguide width (*w*), ridge etching depth (*d*), gap between the signal and ground electrode (*gap*), and width of the SiO$_2$ cladding ($w_{SiO_2}$). A cross-sectional view of the modulator with fabrication parameters is shown in **Figure 2**c.

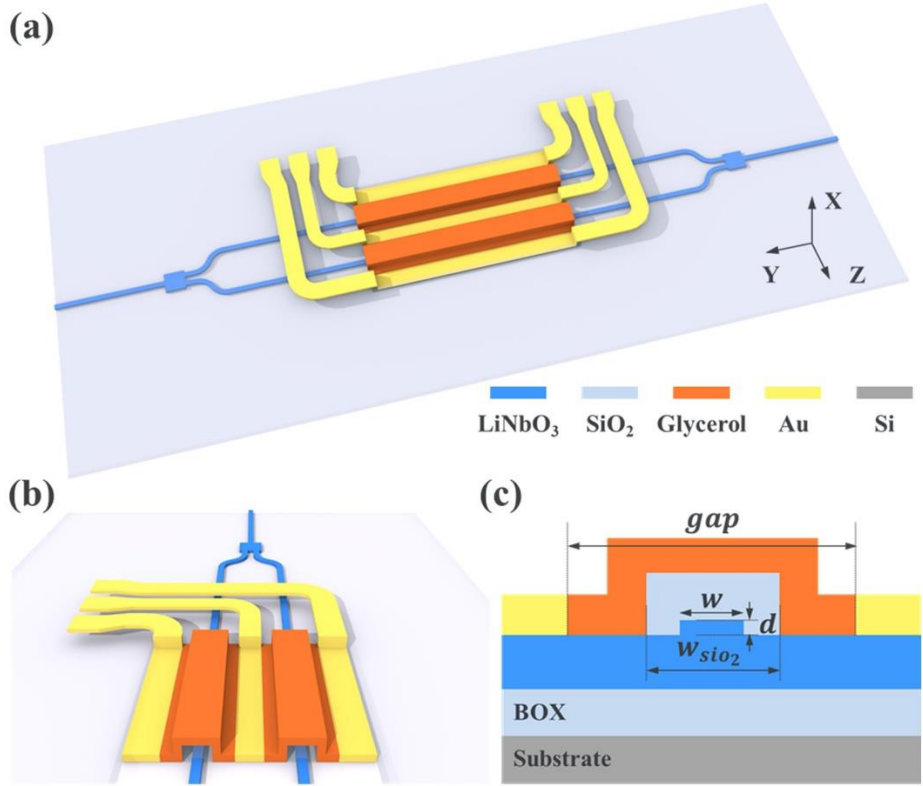

**Figure 2.** a) Three-dimensional schematic of the LNOI modulator. b) Perspective view of the modulator phase shifter. c) Cross-sectional view of the modulator with the fabrication parameters.

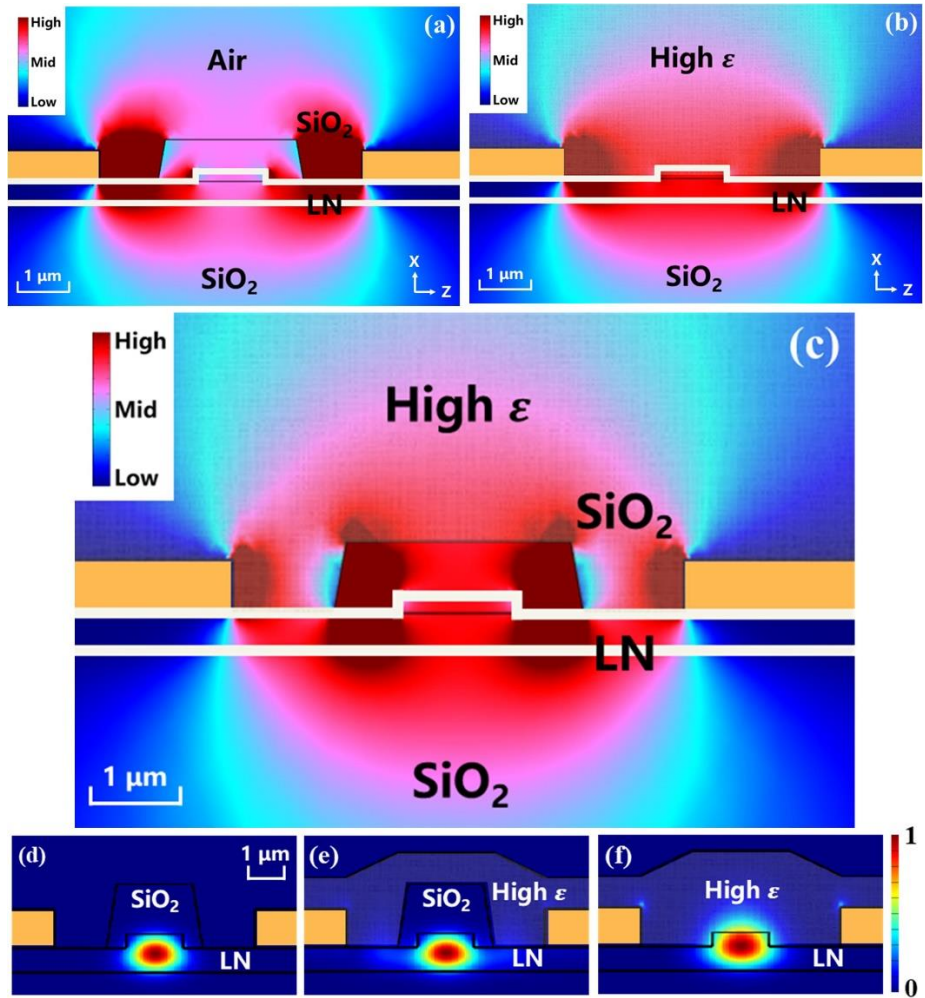

**Figure 3.** Cross-sectional views of the simulated z-component of the microwave field distribution of the TFLN modulator phase shifter with a) the conventional SiO$_2$ cladding, b) pure high-dielectric-constant cladding, and c) composite cladding. Simulated TE optical mode profiles of the phase shifter with d) the conventional SiO$_2$ cladding, e) composite cladding and f) pure high-dielectric-constant cladding, where the optical mode sizes are 1.17 µm$^2$, 2.97 µm$^2$, and 5.57 µm$^2$, respectively.

We simulated the microwave field distribution of a phase shifter with the conventional SiO$_2$ cladding, pure high-dielectric-constant cladding, and composite cladding, as shown in **Figure 3**. For the modulator with the conventional SiO$_2$ cladding, the dielectric constant increases gradually from air to SiO$_2$ ($\varepsilon \sim 3.9$) to LN. Due to the sudden change of the dielectric constant at the interface, the electric field strength is discontinuous at the dielectric level, the ratio being inversely proportional to the dielectric constant of the medium. That is, the electric field is concentrated in the air. However, the electric field strength in the waveguides—which is effective for modulation—is limited.

For the modulator with the pure high-dielectric-constant cladding (glycerol $\varepsilon \sim 44$), because the dielectric constant of glycerol is larger than that of the medium mixed with air and SiO$_2$, the electric field strength inside the LN waveguide is higher than that of the LN waveguide with conventional SiO$_2$ cladding.

For the modulator with the composite cladding, because the dielectric constant of glycerol is larger than that of $SiO_2$, the electric field strength inside the glycerol cladding layer is lower, whereas the electric field strength in the $SiO_2$ cladding layer and the LN ridge waveguides is higher than that of the LN waveguide with conventional $SiO_2$ cladding. Meanwhile, because the electric field distribution inside the LN is continuous, the electric field within the waveguide rises steeply—from low to high—near the interface of the composite dielectric cladding, manifesting as a new electric field intensity peak. The electric field strength in the slab near the ridge waveguide area also increases. Overall, with the combined effect of these two attributes, the electric field strength in LN waveguides is substantially enhanced compared to that of the modulator with the conventional $SiO_2$ cladding and close to that of the modulator with the pure high-dielectric-constant cladding.

However, because high-dielectric-constant materials usually have high refractive indices, they can weaken the optical confinement and reduce the overlap between the electrical and optical fields ($\Gamma$), lowering the electro-optic efficiency. Therefore, we chose to use cladding with a low-refractive-index material. Conversely, we retained the $SiO_2$ cladding to maintain a high-index contrast of >0.7 near the ridge waveguide, keeping strong field confinement.

Because the above-mentioned glycerol has a refractive index of ~1.48—that is, close to that of silica—it cannot adequately reflect the effect of weak optical confinement. Therefore, we used $Si_3N_4$—with its high refractive index of ~1.97—as an example. We simulated the optical field of the transverse electric (TE) mode for the three structures, as shown in **Figure 3**d–f. Composite claddings tightly confine the optical modes compared with pure high-dielectric-constant claddings.

To verify this idea, we used COMSOL Multiphysics and Lumerical Mode software to conduct a co-simulation, considering the effects of the electric and optical field distributions on the modulation efficiency. We simulated the effects of different materials on the modulation efficiency with the same structural parameters. The waveguides featured a top width of $w = 4$ μm and ridge etching depth of $d = 200$ nm. The gap between the signal and ground electrode gap was set to 5 μm, and the width of the $SiO_2$ cladding ($W_{SiO_2}$) was set to 3.2 μm. **Figure 4** shows that the modulation efficiency increases as the relative permittivity ($\varepsilon$) of the upper cladding increases. For the modulators without upper cladding and with glycerol upper cladding, the modulation efficiency increases from 1.95 to 1.56 V·cm, illustrating that composite cladding causes a marked improvement in modulation efficiency.

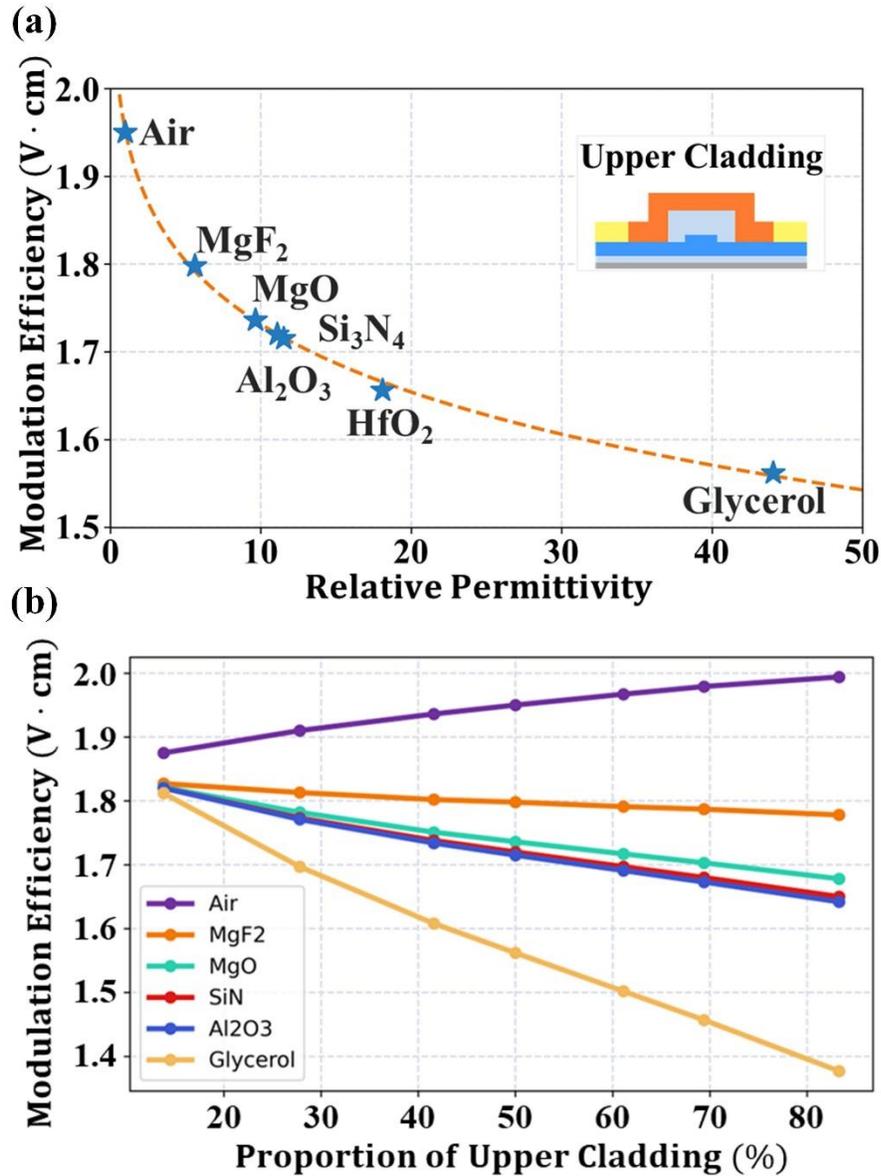

**Figure 4.** a) Simulated modulation efficiencies of modulators with different materials as the upper cladding. b) Simulated modulator efficiencies with different upper claddings versus the proportion of the upper cladding.

To investigate the performance of the proposed design further, we introduced structural variations based on different materials. We used $P = \dfrac{gap - w_{SiO_2}}{gap - w}$ to describe the proportion of the upper cladding and to analyze the effects of $P$ on the modulation efficiency with various materials when the electrode gap (*gap*) was a constant, as shown in **Figure 4**b. When the proportion of the upper cladding increases, the modulator is more linearly efficient. Moreover, when the modulator is almost entirely covered with glycerol, the modulation efficiency is 1.37 V·cm. Air is the only counterexample, as it has a dielectric constant lower than that of $SiO_2$.

## 3. Measurement Results for the Fabricated Device

The proposed LNOI modulators were fabricated and tested. The details of the fabrication and measurement processes are described in the Methods section. **Figure 5** provides microscope and scanning electron microscope (SEM) images of the fabricated device. The right image depicts the cross-sectional view of the modulator phase shifter. The composite cladding consists of 800-nm-thick $SiO_2$ and 500-nm-thick high-dielectric-constant material.

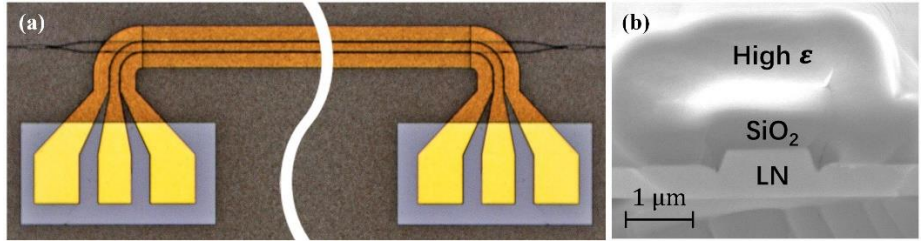

**Figure 5.** a) Microscope image of the fabricated LNOI modulator. b) SEM image of the cross-sectional view of the modulator phase shifter.

We characterized the modulation efficiency of the fabricated LNOI modulator using the experimental setup shown in **Figure 6**b. More than five materials were utilized as the upper cladding to confirm the reliability and universality of our study. However, owing to process limitations, some cladding films fractured, resulting in vacant data. **Figure 6**d shows the EO half-wave voltage measurement for the 4-mm-long devices with different materials as the upper cladding when the width of the $SiO_2$ cladding $w_{SiO2} = 3.2$ μm. The experimental results agree well with the simulation results.

The modulation efficiency increases as the relative permittivity ($\varepsilon$) of the upper cladding increases. **Figure 6**e shows the influence of the proportion of the upper cladding on the modulation efficiency. The experimental results follow the rule that when the proportion of the upper cladding increases, the modulator becomes more efficient. We found the measured modulation efficiency using high-refractive-index cladding to be worse than that in the simulation, which may be due to the weak optical confinement and the deterioration of the overlap factor being more than expected. When $P \approx 85\%$, there is a reverse-trend data point, which could be a result of an overlay error—the finer the structure, the more it is affected. The impact of the upper cladding leads to more light leakage, which influences the modulation efficiency.

Overall, there is good agreement between the simulated and measured data, proving that the simulation offers useful guidance for improving modulator performance. Glycerol is the most suitable cladding material, satisfying the need for both a low refractive index and a high dielectric constant. When the modulator is almost entirely covered with glycerol, the measured $V\pi$ is 3.52 V, and the corresponding voltage–length product ($V\pi L$) is 1.41 V·cm.

Additionally, we characterized the EO response of our modulator using a 4-mm-length phase shifter. As shown in **Figure 6**f, the trend of the curve for the 3 dB bandwidth is higher than the (40 GHz) range of our measurement system. This result indicates that the structure of the composite cladding will not affect the high-frequency performance of the modulators.

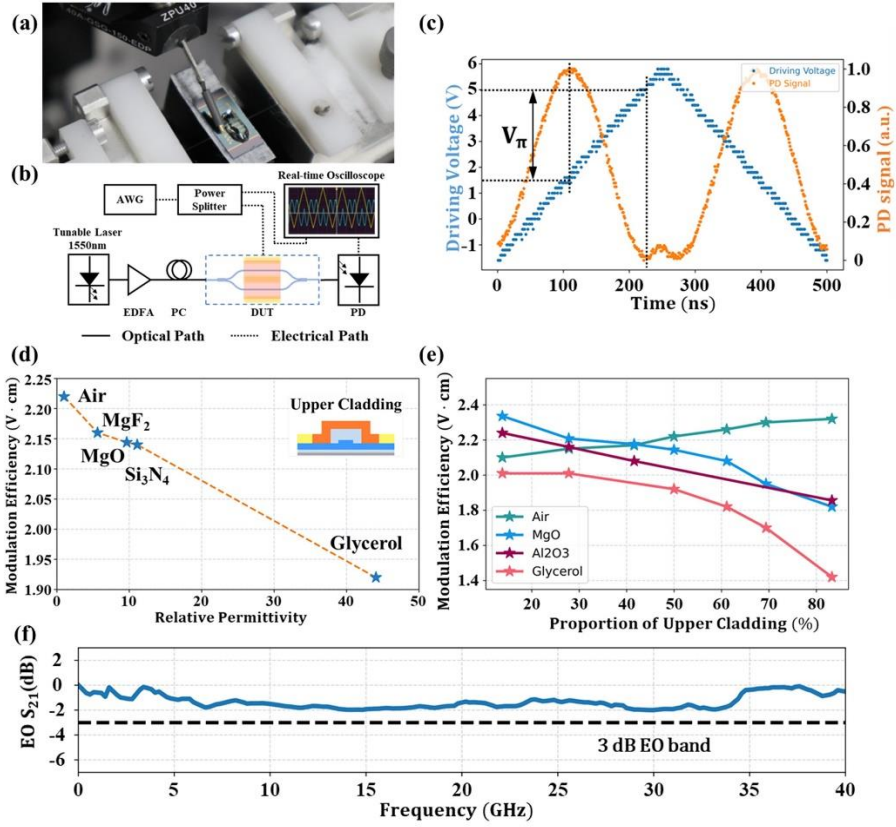

**Figure 6.** a) Microscope image of the device under test. b) Setup for modulation efficiency measurement. c) Normalized optical transmission of the 4 mm device as a function of the applied voltage, showing Vπ of 3.52 V and VπL of 1.41 V·cm. d) Measured modulation efficiencies of modulators with different materials as the upper cladding. e) Measured modulation efficiencies of modulators with different materials as the upper cladding versus the proportion of the upper cladding. f) EO response of the 4-mm-long device. The 3 dB bandwidths of both devices are beyond the measurement limit of the VNA (40 GHz).

The performances of the reported LNOI MZI modulators are summarized in **Table 1**. Among these modulators, our modulator fabricated via standard photolithography shows lowest voltage-length products and the lowest optical loss for a broadband lithium niobate modulator so far. Our results are significantly superior to those in reference [29] in terms of both modulation efficiency and loss, while having no impact on the device's high-frequency performance in comparison to reference [30]. In brief, the modulator developed in this study achieved a satisfactory compromise among modulation efficiency, loss, and bandwidth.

**Table 1. Performance comparison of the reported LNOI MZI modulators**

| Ref. | Device length (mm) | $V_\pi \cdot L$ (V·cm) | Loss (dB) | 3 dB EO bandwidth (GHz) |
|---|---|---|---|---|
| [7] | 5 | 2.2 | 0.5 | 100 GHz |
| [20] | 5 | 6.7 | - | 106 GHz |
| [21] | 24 | 3.7 | 5.4 | 29 GHz |
| [22] | 5 | 2.5 | 2.5 | 70 GHz |
| [29] | 5 | 1.75 | 2.5 | >40 GHz |
| [30] | 7.5 | 1.32 | 2 | >3 GHz |
| This Work | 4 | 1.41 | 0.5 | >40 GHz |

## 4. Conclusions

We experimentally demonstrated an LNOI high-efficiency EO modulator with composite cladding. The structure of the composite cladding could increase the electric field strength in the LNOI waveguide, thereby improving the modulation efficiency without negatively impacting the high-frequency performance. The measured voltage–length product was 1.41 V·cm when using glycerol cladding. This approach is a novel method of improving the modulation efficiency based on the traditional MZI modulator structure, which could be combined with other structures to achieve better performance. Our suggested design strategy of using high-permittivity cladding may provide a means of enhancing photonic device integration on the TFLN platform and function as a key element in future large-scale photonic integrated circuits.

## 5. Experimental Section/Methods

### 5.1. Device Fabrication

The devices were fabricated on an LNOI wafer from NANOLN using a 600-nm-thick X-cut TFLN layer and a 4.7-µm-thick buried $SiO_2$ layer. The pattern was defined using an i-line stepper, and a 200-nm-thick LN layer was etched using inductively-coupled plasma (ICP) reactive-ion etching. Subsequently, an 800-nm-thick $SiO_2$ layer was deposited on the top of the waveguides using plasma-enhanced chemical vapor deposition and etched using ICP. A 600-nm-thick Ti/Au layer was subsequently deposited via electron-beam evaporation and removed to form the electrodes. Finally, a 500-nm-thick high-dielectric-constant material layer was deposited using electron-beam evaporation and removed to leave the probing spaces.

### 5.2. Device Characterization

The experimental setup is shown in **Figure 6**b. A 1550 nm light from a tunable laser (SANTEC TSL-550) was amplified using an erbium-doped fiber amplifier (EDFA: Keopsys CEFA-C-PB) and subsequently launched into the device under test. A polarization controller was used to adjust the input light to TE mode. Subsequently, the modulated light was detected using a high-speed photodetector. An arbitrary signal generator (AWG: GW Instek AFG-3051) was employed to provide a 1 MHz triangular voltage sweep. A real-time oscilloscope was used to measure the signal from the AWG and the photodetector (PD: LSIHPD-A12G). To quantify the system loss, reference waveguides were attached to the chip. As a result, the fiber–waveguide coupling loss and straight waveguide loss could be filtered to derive the normalized on-chip insertion loss of the device.

## Acknowledgements

This work was supported by National Key R&D Program of China(2019YFB2203702). The authors acknowledge Zhejiang University Micro and Nano Processing Platform for providing the facility support. The authors thank Mrs. Qingjiao Mi for assistance with the experiments.

## Conflict of Interest

The authors declare no conflict of interest.

## Data Availability Statement

The data that support the findings of this study are available from the corresponding author upon reasonable request.